\documentclass[12pt]{article}
\usepackage{amssymb,amsmath}
\usepackage[noblocks]{authblk}
\usepackage{graphicx}
\usepackage{epstopdf}
\usepackage[top=0.75in, bottom=0.75in, left=0.75in, right=0.75in, dvips]{geometry}
\usepackage{caption}
\begin{document}
\textwidth 10.0in
\textheight 9.0in
\topmargin -0.60in
\title{Renormalization Scheme Dependence\\
 in a QCD Cross Section}
\author[1]{Farrukh Chishtie}
\author[2,3]{D.G.C. McKeon}
\author[4]{T.N. Sherry}
\affil[1] {Theoretical Research Institute, Pakistan Academy of Sciences (TRIPAS)\\
Islamabad, Pakistan 44000}
\affil[2] {Department of Applied Mathematics, The
University of Western Ontario, London, ON N6A 5B7, Canada}
\affil[3] {Department of Mathematics and
Computer Science, Algoma University,\newline Sault Ste. Marie, ON P6A
2G4, Canada}
\affil[4] {School of Mathematics, Statistics and Applied Mathematics, National University of Ireland Galway, University Road, Galway, Ireland} 
\maketitle

\maketitle
email: fchishti@uwo.ca, dgmckeo2@uwo.ca

\begin{abstract}
The zero to four loop contribution to the cross section $R_{e^{+}e^{-}}$ for $e^{+}e^{-} \longrightarrow$ hadrons, when combined with the renormalization group equation, allows for summation of all leading-log ($LL$), next-to-leading-log $(NLL) \ldots N^3LL$ perturbative contributions.  

It is shown how all logarithmic contributions to $R_{e^{+}e^{-}}$ can be summed and that $R_{e^{+}e^{-}}$ can be expressed in terms of the log independent contributions, and once this is done the running coupling $a$ is evaluated at a point independent of the renormalization scale $\mu$.  All explicit dependence of $R_{e^{+}e^{-}}$ on $\mu$ cancels against its implicit dependence on $\mu$ through the running coupling $a$ so that the ambiguity associated with the value of $\mu$ is shown to disappear. The renormalization scheme dependency of the ``summed'' cross section $R_{e^{+}e^{-}}$ is examined in three distinct renormalization schemes.  In the first two schemes, $R_{e^{+}e^{-}}$ is expressible in terms of renormalization scheme independent parameters $\tau_i$ and is  explicitly and implicitly independent of the renormalization scale $\mu$.

Two of the forms are then compared graphically both with each other and with the purely perturbative results and the $RG$-summed $N^3LL$ results.
\end{abstract}

\section{Introduction}

With the discovery of asymptotic freedom, perturbative calculations in QCD became feasible.  In QED, the renormalization scheme ambiguities that arise in the course of loop calculations can generally be ignored as there appears to be a ``natural'' renormalization scheme which minimizes higher order effects.  In QCD however, no such scheme exists and by varying renormalization scheme, one can widely vary the results of higher loop calculations.

Technically, the easiest renormalization schemes to implement are the mass-independent ones [1,2]; at higher loop order poles in $\epsilon = 2 - d/2$ ($d$-number of dimensions) can be removed through ``minimal subtraction'' [2,3] when combined with the BPHZ renormalization procedure [4]. As a result of this procedure, the form of the cross section $R_{e^{+}e^{-}}$ is given by $3(\sum_i q_i^2)(1+R)$ where $R$ at $n$-loop order has a perturbative contribution of order $a^{n+1}$ in 
%1
\begin{equation}
R = R_{\rm{pert}} = \sum\limits_{n=0}^\infty r_n a^{n+1} = \sum\limits_{n=0}^\infty \sum\limits_{m=0}^n \, T_{n,m} L^m a^{n+1}\qquad (T_{0,0} = 1)
\end{equation}
with $L = b \ln (\frac{\mu}{Q}), Q^2$ being the centre of mass momentum squared. (The constant $b$ also occurs in eq. (2) below.)

The explicit dependence of $R$ on the renormalization scale parameter $\mu$ is compensated for by implicit dependence of the ``running coupling'' $a(\mu)$ on $\mu$,
\begin{equation}
 \mu \frac{\partial a}{\partial \mu} = \beta (a) = -ba^2 \left( 1 + ca + c_2a^2 + \ldots\right).
\end{equation}
(Our definition of $a(\mu)$ coincides with that of ref. [13].)
In general, when using mass-independent renormalization, different renormalizations schemes have their couplings $a$ and $\overline{a}$ related by [5]
%3
\begin{subequations}
\begin{align}
\overline{a} &= a + x_2a^2 + x_3a^3 + \ldots\\
& \equiv F(a) \nonumber
\intertext{and it follows that $b$ and $c$ in eq. (2) are scheme independent while the $c_n(n \geq 2)$ are scheme dependent.
From the equation $\overline{\beta}(\overline{a}) = \beta (a) F^\prime (a)$, it follows that [44]}
\overline{c}_2 &= c_2 - c x_2 + x_3 - x_2^2 \\
\overline{c}_3 &= c_3 -3 c x_2^2 + 2(c_2 - 2\overline{c}_2) x_2 + 2x_4 - 2x_2x_3
\end{align}
\end{subequations} 
etc.

Various strategies have been used both to minimize the dependence of perturbative results in QCD on both $\mu$ and on general scheme dependency.  (If the exact result for $R$ were known, then all such dependency should disappear [6].)  One option is to choose a ``physical'' value for $\mu$, such as setting $\mu = Q$.  In this case, $L = 0$ and only $T_{n,0}$ contributes to the sum in eq. (1); all $Q$ dependence now resides in $a(Q/\Lambda)$ where $\Lambda$ is a dimensionful constant associated with the boundary condition imposed on eq. (2).  The $Q$ dependency of $a$ is then interpreted to be the result of ``summing the logs'' in eq. (1).  We will demonstrate below a more concrete way of summing logs in eq. (1) through use of the renormalization group (RG) equation [34-36].

Quite often, the mass independent renormalization schemes are viewed as being so ``unphysical'' that results derived through their use cannot be trusted; the contributions coming from loop effects beyond the order to which calculations have been performed are taken to be non-negligible.  More ``physical'' renormalization schemes, such as ``momentum subtraction'' ($MOM$) [7,8,9] or the ``$V$-scheme'' which relies on calculating the effective potential between two heavy quarks [10,11,12] are taken to generate perturbative results that are better approximations to the exact result than these derived using $\overline{MS}$.  We note though that these renormalization schemes are more difficult to implement and also result in additional ambiguities.  For example, even with the $MOM$ schemes for renormalization, in eq. (2) not only are the $c_n(n \geq 2)$ scheme dependent, but so is $c$ itself.  In addition both $c$ and $c_n (n \geq 2)$ are gauge dependent with the Landau gauge being ``preferable'' as in this gauge both $b$ and $c$ in eq. (2) coincide with values obtained using  $\overline{MS}$.

When one uses  $\overline{MS}$, a variety of strategies have been adopted to minimize the renormalization scheme dependence in the sum of eq. (1).  When using the ``principle of minimal sensitivity'' ($PMS$) the parameters $\mu$ and $c_i$ are chosen so that the variations of $R_{e^{+}e^{-}}$ when these parameters are altered is minimized [13].  In the ``fastest apparent convergence'' ($FAC$) approach, contributions beyond a given order in perturbation theory are minimized by the introduction of ``effective charges'' [14,15,16,17].  Another method involves the ``principle of maximum conformality'' ($PMC$) in which a different renormalization mass scale is introduced at each order of perturbation theory; these mass scales are then chosen to absorb all dependence on the coefficients $c_i$ occurring in eq. (2) at the order of perturbation theory being considered [18,19,20]. This approach has been extensively
developed in ref. [50].

One can also simply employ ``renormalization group summation'' ($RG\sum$).  In this approach the $RG$ equation with one loop $RG$ functions permits summation of all ``leading-log'' ($LL$) contributions to the sum in eq. (1), two loop $RG$ functions permits summation of all ``next-to-leading-log'' ($NLL$) contributions etc.  This has been applied to a number of perturbative calculations in $QCD$ [21,22,23] as well as thermal field theory, the effective action [24,25] etc.  As expected, $RG\sum$ reduces the dependence of any calculation on the scale parameter $\mu$, which one might anticipate as upon including higher order logarithmic effects, one should be closer to the exact result, which is fully independent of $\mu$.  Of course any computation to finite order in perturbation theory is scheme dependent.

There is another way of organizing the sum of eq. (1).  Instead of computing the $LL$, $NLL$ etc. sums in turn, one can use the $RG$ equation to show that all logarithmic contributions to $R$ can be expressed in terms of the log-independent contributions.  We will show that by using this summation, the explicit dependence of $R_{e^{+}e^{-}}$ on $\mu$ occurring in eq. (1) through $L$ is exactly cancelled by the implicit dependence on $\mu$ through the running coupling $a(\mu)$ [32].

Three particular mass independent renormalization schemes are then considered; in each of them $R$ is expressed in terms of renormalization scheme invariants $\tau_i$ after making a convenient choice of the parameters $c_i$ in eq. (2).  In one of these schemes, $T_{n,0} = 0 (n \geq 2)$ while in the other two schemes, the first has $c_i = 0 (i \geq 2)$ (the ``'t Hooft scheme'' [26]) while the other has $c_i = 0 (i \geq 3)$.  Some features of these three schemes are examined.  This involves deriving explicit series showing how a change of $\mu$ and a change of $c_i$ affects the running coupling.

We also illustrate graphically some features of this approach to computing $R$ in terms of renormalization scheme invariants; the graphical analysis demonstrates  improvements over both the strictly perturbative results and those obtained using $RG\Sigma$ alone.

\section{Renormalization Group Summation}

In order to sum $LL$, $NLL$ etc. contributions to $R$ in eq. (1) we use the groupings
\begin{equation}
S_n(aL) = \sum\limits_{k=0}^\infty \,\,T_{n+k,k} (aL)^k
\end{equation}
so that the $RG$ equation
\begin{equation}
\left( \mu \frac{\partial}{\partial\mu} + \beta(a) \frac{\partial}{\partial a} \right) R = 0
\end{equation}
with $\beta(a)$ given by eq. (2) leads to a set of nested equations with the boundary conditions
\begin{equation}
S_n(0) = T_{n,0} \equiv T_n
\end{equation}
Using eq. (2), we find that eq. (5) is satisfied order by order in $a$ provided $S_n(u)$ satisfies
\begin{subequations}
\begin{align}
&S_0^\prime - (S_0 + u S_0^\prime) = 0\\
&S_1^\prime - (2S_1 + u S_1^\prime) - c (S_0 +uS_0^\prime) = 0\\
&S_2^\prime - (3 S_2 + uS_2^\prime) - c(2S_1 + u S_1^\prime) - c_2 (S_0 + u S_0^\prime) = 0\\
&S_3^\prime - (4 S_3 + uS_3^\prime) - c(3S_2 + u S_2^\prime) -c_2(2S_1 + u S_1^\prime)
- c_3 (S_0 + u S_0^\prime) = 0\\
& \mathrm{etc.}\nonumber
\end{align}
\end{subequations}
so that [22,23]
\begin{subequations}
\begin{align}
wS_0 &= T_{00} \quad (w = 1 - u)\\
w^2S_1 &=T_{10} - c T_{00} \ln |w|\\
w^3 S_2 &= T_{20} - (2c T_{10} + c^2 T_{00})\ln |w| + (c^2 - c_2) T_{00} (w-1) + c^2T_{00} \ln^2 |w|\\
w^4S_4 &= T_{30} - c^3 T_{00}\ln^3 |w| + \frac{1}{2}(6c^2 T_{10} + 5c^3 T_{00})  \ln^2 |w| -2c (c^2 - c_2) T_{00}(w\ln |w| - (w-1)) \nonumber \\
&- 3c(T_{20} - (c^2-c_2) T_{00})\ln |w| + (-2c_2 T_{10} -c(2c^2 -c_2)T_{00})(w-1)+ (-c^3 + 2cc_2 - c_3)T_{00}(\frac{w^2-1}{2})\\
& \mathrm{etc.}\nonumber
\end{align}
\end{subequations}
where the $S_i(i = 0,1\ldots 4)$ are the $LL$, $NLL$, $N^2LL$ and $N^3LL$  contributions to $R$.

If instead of using
\begin{equation}
R = R_\Sigma = \sum_{n=0}^\infty a^{n+1} S_n(aL)
\end{equation}
we directly substituted eq. (1) into eq. (5) we find that
\begin{subequations}
\begin{align}
T_{ii} &= T_{i-1,i-1} \\
T_{21} &= (c + 2 T_{10})\\
2T_{32}& = (2cT_{11} + 3T_{21})\\
\intertext{and}
T_{31} &= c_2 + 3T_{20} + 2cT_{10}
\end{align}
\end{subequations}
etc.

Instead of the grouping of eq. (4) we can also set
\begin{equation}
A_n = \sum\limits_{k=0}^\infty \, T_{n+k,n} a^{n+k+1}
\end{equation}
so that now
\begin{equation}
R = R_A = \sum_{n=0}^\infty A_n(a)L^n.
\end{equation}
Eq. (5) is satisfied if
\begin{equation}
A_n(a) = - \frac{\beta(a)}{bn} \frac{d}{da} A_{ n-1}(a) .
\end{equation}
We now define $\eta = \ln \frac{\mu}{\Lambda}$ where $\Lambda$ is a universal scale associated with the boundary condition on eq. (2) so that

\begin{equation}
\eta = \int_{a_{I}}^{a(\eta)} \frac{dx}{\beta(x)} \quad (a_I = a(\eta = 0) = \mathrm{const.}).
\end{equation}
By eqs. (2,13) we find that
\begin{equation}
A_n(a(\eta)) = \frac{-1}{bn}\frac{d}{d\eta} A_{n-1} (a(\eta)) = \frac{1}{n!}\left( - \frac{1}{b}\frac{d}{d\eta}\right)^n A_0 (a(\eta)).
\end{equation}
Together, eqs. (12,15) lead to
\begin{equation}
 R_A  = \sum_{n=0}^\infty \frac{1}{n!} \left( - \frac{L}{b}\right)^n \frac{d^n}{d\eta^n} A_0 (a(\eta))
\end{equation}
\begin{equation}
= A_0 \left( a\left( \eta - \frac{1}{b} L\right)\right).
\end{equation}
With the definitions of $\eta$ and $L$, we see that eq. (17) becomes
\begin{equation}
R_A = A_0 \left( a \left(\ln \frac{Q}{\Lambda}\right)\right).
\end{equation}
This is an exact equation that expresses $R$ in terms of its log independent contributions and the running coupling $a$ evaluated at $\ln \frac{Q}{\Lambda}$ with all dependence of $R$ on $\mu$, both implicit and explicit, removed.  This disappearance of dependence on $\mu$ is to be expected as $\mu$ is unphysical.  The mass parameter $\Lambda$ is a reference point at which the value of $a(\eta)$ is measured experimentally; when $Q = \Lambda$ then $\ln Q/\Lambda = \eta = 0$ and $a(\eta = 0) = a_I$.  $\Lambda$ is defined in refs. [8,13] by the equation
\begin{equation}
\ln \left( \frac{Q}{\Lambda}\right) = \int_0^{a\left(\ln \frac{Q}{\Lambda}\right)}
\frac{dx}{\beta(x)} + \int_0^\infty \frac{dx}{bx^2(1 + cx)}
\end{equation}
so that with this choice of defining $\Lambda$ (by eqs. (14,19)) we find that $a_I$ is given by
\begin{equation}
0 = \int_0^{a_I} \frac{dx}{\beta(x)} + \int_0^\infty \frac{dx}{bx^2(1+c x)} .
\end{equation}
It is clear that both $\Lambda$ and $a_I$ are renormalization scheme dependent; under a change of renormalization scheme given in eq. (3), $\Lambda$ in eq. (19) is affected by $x_2$ [8,13] so that
\begin{equation}
\frac{\overline{\Lambda}}{\Lambda} = \exp \left( \frac{x_2}{b}\right).
\end{equation}

We can see how $a(\mu)$ evolves under a charge of $\mu$ either by direct examination of eq. (2) or by use of the series expansion
\begin{align}
 a_j &= a_i + \left(\sigma_{21} \ell \right) a_i^2 + \left(\sigma_{31} \ell + \sigma_{32} \ell^2 \right)a_i^3 \\
 &+ \left(\sigma_{41} \ell + \sigma_{42} \ell^2 +\sigma_{43} \ell^3 \right) a_i^4 + \ldots \nonumber 
\end{align}
where $a_i \equiv a(\mu_i)$ and $\ell = \ln\left(\frac{\mu_i}{\mu_j}\right)$.  Since $a_j$ is independent of $\mu_i$, $\mu_i \frac{d}{d\mu_i} a_j = 0$ and thus by eq. (2)
%24
\begin{align}
0 = \beta(a) \bigg[ 1 &+ 2(\sigma_{21}\ell ) a_i + 3 (\sigma_{31}\ell + \sigma_{32} \ell^2)a_i^2 \\
&+ 4(\sigma_{41}\ell + \sigma_{42} \ell^2 + \sigma_{43} \ell^3) a_i^3 + \ldots \bigg]\nonumber \\
&+ \bigg[ (\sigma_{21}) a_i^2 + (\sigma_{31} + 2\sigma_{32} \ell) a_i^3 +
(\sigma_{41} + 2\sigma_{42} \ell + 3\sigma_{43} \ell^2) a_2^4 + \ldots \bigg]\nonumber
\end{align}
from which we obtain
\begin{equation}
 a_j = a_i + b \ell a_i^2 + \left( bc \ell +b^2 \ell^2 \right) a_i^3 + 
\left( bc_2\ell  + \frac{5}{2} b^2c \ell^2 + b^3 \ell^3\right) a_i^4  + \ldots \;. 
\end{equation}
Eq. (24) relates the value of the running coupling $a(\mu)$ at different values of $\mu$.  It follows from eq. (24) that
\begin{align}
a\left( \ln \frac{Q}{\Lambda}\right) = a_I + b\left( \ln \frac{Q}{\Lambda}\right)  & a_I^2 + \left(bc \left( \ln \frac{Q}{\Lambda}\right) + b^2 \left( \ln \frac{Q}{\Lambda}\right)^2\right) a_I^3 \\
& + \left(bc_2  \ln \frac{Q}{\Lambda} + \frac{5}{2} b^2c \left( \ln \frac{Q}{\Lambda}\right)^2 + b^3 \left( \ln \frac{Q}{\Lambda}\right)^3\right) a_I^4 + \ldots \nonumber
\end{align}

We now examine the renormalization scheme dependency of eq. (18).

\section{Renormalization Scheme Dependence of $R_A$}

As noted in the introduction, the coefficients $c_i (i \geq 2)$ in eq. (2) are renormalization scheme dependent; indeed Stevenson [13] has shown that these parameters can be used to characterize the renormalization scheme being used when using mass independent renormalization. If we have
\begin{equation}
\frac{\partial a}{\partial c_i} = \beta_i(a) 
\end{equation}
then it follows from 
\begin{equation}
\mu \frac{\partial^2 a}{\partial\mu \partial c_i^2} -\mu \frac{\partial^2 a}{\partial c_i\partial\mu} = 0 
\end{equation}
that
\begin{align}
\beta_i(a)&= -b\beta(a)  \int_0^a dx \frac{x^{i+2}}{\beta^2(x)}\\
& \approx \frac{a^{i+1}}{i-1} \bigg[ 1 + \left( \frac{(-i+2)c}{i}\right)a + \left( \frac{(i^2-3i+2)c^2+(-i^2+3i)c_2}{(i+1)i}\right)a^2\nonumber \\
& + \left( \frac{(-i^3+3i^2+4i)c_3 + (2i^3 -6i^2 +4)cc_2+ (-i^3+3i^2-2i)c^3}{(i+2)(i+1)i}\right)a^3 + \ldots \bigg].
\end{align}

As $R_A$ in eq. (18) is independent of the choice of renormalization scheme, it follows that
\begin{equation}
\left( \frac{\partial}{\partial c_i} + \beta_i (a) \frac{\partial}{\partial a} \right) R_A \left( a \left(\ln \frac{Q}{\Lambda}\right)\right) = 0.
\end{equation}
With $A_0$ defined by eq. (11) and $\beta_i(a)$ given in eq. (29), it follows from eq. (30) that $T_{n0} \equiv T_n$ obey a set of differential equations whose solutions are 
\begin{align}
T_0 = \tau_0 = 1,& \quad T_1 = \tau_1, \quad T_2 = -c_2 + \tau_2,\quad T_3 = -2c_2\tau_1 - \frac{1}{2} c_3 + \tau_3\nonumber \\
&T_4 = -\frac{1}{3} c_4 - \frac{c_3}{2} \left( - \frac{1}{3} c+2\tau_1\right) + \frac{4}{3} c_2^2 - 3c_2\tau_2 + \tau_4\nonumber \\
&T_5 = \left[ \frac{1}{3} cc_2^2 + \frac{3}{2}c_2c_3 +\frac{11}{3} c_2^2 \tau_1  -  4c_2  \tau_3\right]\nonumber\\
& \quad -\frac{1}{2}\left[ \frac{1}{6} c^2c_3 - \frac{2}{3} c_3c \tau_1 + 3c_3\tau_2\right]\nonumber \\
& \quad -\frac{1}{3}\left[ -\frac{1}{2} c_4c + \frac{1}{2} c_4 \tau_1\right] - \frac{1}{4} c_5 + \tau_5 \nonumber \\
& \intertext{etc.}\nonumber
\end{align}
The constants of integration $\tau_i$ appearing in eq. (31) are renormalization scheme invariants.

We now relate couplings $a_c$ and $a_d$ associated respectively with the parameters $c_i$ and $d_i$  by the expansion
\begin{equation}
a_{c} = a_d + \lambda_2 (c_i, d_i) a^2_d + \lambda_3 (c_i, d_i) a_d^3  + \ldots
\end{equation}
where $\lambda_N (c_i, c_i) = 0$ and $\lambda_{N-1}$ can depend on $d_2 \ldots d_N$. In eq. (32), $\beta_c(a_c)$ is given by eq. (2) and $\beta_d(a_d) = - ba_d^2(1 + ca_d + d_2 a_d^2 + \ldots )$.  Since $\frac{d a_c}{dd_i} = 0$, we find that 
\begin{equation}
\left( \frac{\partial}{\partial d_j} + \beta_j (a_d) \frac{\partial}{\partial a_d}\right) \sum_{N=1}^\infty \lambda_N (c_i, d_i) a^N_d = 0 
\end{equation}
from which we obtain a set of differential equations for $\lambda_N$; their solution leads to 
\begin{align}
a_c = a_d & - (d_2 - c_2)  a^3_d - \frac{1}{2} (d_3 - c_3) a^4_d \nonumber \\
&+ \left[ - \frac{1}{6} \left( d_2^2 - c_2^2\right) + \frac{3}{2} \left(d_2 - c_2\right)^2 + \frac{c}{6} \left(d_3 - c_3 \right)
 - \frac{1}{3} \left(d_4 - c_4 \right) \right] a^5_d + \ldots .
\end{align}
Eq. (34) satisfies the group property so that $a_c(a_d(a_e)) = a_c(a_e)$.

We now will consider three specific choices of renormalization scheme and what effect such choices have on the form of $R_A$. (The running coupling when using the $I^{th}$ scheme is denoted by $a_{(I)}$.)  The first scheme involves a very particular choice of $c_i(i \geq 2)$ chosen so that $T_n = 0 (n \geq 2)$.  From eq. (31) we find
\begin{align}
c_2 &= \tau_2\\
c_3 &= -4 \tau_2\tau_1 + 2\tau_3\nonumber \\
c_4 &=  c(\tau_3 - 2 \tau_1\tau_2) + 12 \tau_1^2 \tau_2 - 6 \tau_1\tau_3 - 5 \tau_2^2 + 3\tau_4 \nonumber\\
c_5 &=  \left[ \frac{4}{3} c\tau_2^2 + \frac{44}{3} \tau_2^2\tau_1 - 16 \tau_2\tau_3 \right]\nonumber \\
&\quad+ \left[2\tau_3- 4 \tau_1\tau_2 \right]
 \left[ 6\tau_2 - \frac{1}{3} c^2 + \frac{4}{3} c\tau_1 - 6 \tau_2 \right]\nonumber\\
&\quad +\left[c(\tau_3 - 2 \tau_1\tau_2) + 12 \tau_1^2\tau_2 - 6 \tau_1\tau_3 - 5 \tau_2^2 + 3\tau_4 \right]\left[ \frac{2}{3} (c - \tau_1)\right] + 4\tau_5 .\nonumber
 \end{align}
In this first case, we find that $R_A$ in eq. (18) reduces to just the finite sum
\begin{equation}
R = R_A^{(1)} = a_{(1)} \left( \ln \frac{Q}{\Lambda} \right) + \tau_1 a_{(1)}^2 \left(\ln \frac{Q}{\Lambda} \right).
\end{equation}

A second choice of renormalization scheme is to take $c_i = 0 (i \geq 2)$.  With this choice, $T_n = \tau_n$ and so by eq. (18) 
\begin{equation}
R = R_A^{(2)} = a_{(2)} \left( \ln \frac{Q}{\Lambda} \right) + \tau_1 a_{(2)}^2
\left( \ln \frac{Q}{\Lambda} \right) + \tau_2 a_{(2)}^3 \left( \ln \frac{Q}{\Lambda} \right) + \ldots
\end{equation}
where $a_{(2)} \left( \ln \frac{Q}{\Lambda} \right)$ is given by 
\begin{equation}
 \ln \left(\frac{Q}{\Lambda} \right) = \int_{a_{I(2)}}^{a_{(2)}\left( \ln \frac{Q}{\Lambda} \right)} \frac{dx}{-bx^2(1 + cx)} .
\end{equation}
It is well known that this integral can be done in closed form, leading to $a_{(2)}$ being expressed in terms of the Lambert $W$ function [24,27,28].

Since $R_A^{(1)} = R_A^{(2)}$, we find that 
\begin{equation}
a_{(1)} + \tau_1 a_{(1)}^2 = a_{(2)} + \tau_1 a_{(2)}^2 +
\tau_2 a_{(2)}^3 + \tau_3 a_{(2)}^4 + \ldots
\end{equation}
which is consistent with eq. (34).

A third choice would be to take $c_i = 0 ( i \geq 3)$ so that by eq. (31)
\begin{align}
R = R_A^{(3)} = a_{(3)}\left( \ln \frac{Q}{\Lambda} \right)& + \tau_1 a_{(3)}^2\left( \ln \frac{Q}{\Lambda} \right) + (\tau_2 - c_2) a_{(3)}^3\left( \ln \frac{Q}{\Lambda} \right)\\
& + (\tau_3-2c_2\tau_1) a_{(3)}^4\left( \ln \frac{Q}{\Lambda} \right) +
(\tau_4 - 3c_2\tau_2 + \frac{4}{3} c_2^2) a_{(3)}^5\left( \ln \frac{Q}{\Lambda} \right) + \ldots \nonumber
\end{align}
With this choice of $c_i$, $\beta (a) = - ba^2 (1 + ca + c_2a^2)$ and eq. (14) reduces to
\begin{equation}
 \ln \frac{Q}{\Lambda} = \int_{a_{I(3)}}^{a_{(3)}\left( \ln \frac{Q}{\Lambda} \right)} \frac{dx}{-bx^2(1 + c x + c_2x^2)}
\end{equation}
with $c_2$ unspecified.

We can now compare the different approaches to computing $R$.

\section{Comparing Computations}

We now will discuss how the considerations in the preceding sections can be used in conjunction with explicit perturbative calculations that have been performed to analyze experiments.  In particular we will see how the renormalization schemes which lead to $R_A^{(1)}$ (eq. (36)) and $R_A^{(2)}$ (eq. (37)) can be compared with experimental results.

When using the $\overline{MS}$ scheme, explicit calculation of the $\beta$ function has been done to four loop order [29]; $R$ has also been computed to four loop order using this scheme [30]. The five-loop $O(\alpha_s^4)$ corrections to $R$ were estimated by
different methods, including those using the Asymptotic Pade Approximation Procedure (APAP) [46] and were analytically evaluated in [37]. The five loop contribution to the beta function has been discussed in ref. [47] and computed analytically in the MSbar scheme [48]. (See also ref. [49].) Here we restrict
ourselves to the four loop calculations of $R$ and $\beta$ which allow us to determine the renormalization scheme invariants $\tau_1$, $\tau_2$, $\tau_3$ appropriate for this process by using eq. (31).  In addition, we can use the measured value of $a$ in the $\overline{MS}$ scheme at some mass scale (in particular, the $\tau$ lepton mass $m_\tau$) and then use this value to evaluate $a$ at a different mass scale (by using eq. (24)) and in a different renormalization scheme (by using eq. (34)).

We have made four plots which are collated in fig. 1 to illustrate and compare results from the various ways four loop calculations can be used to determine $R$ in eq. (1).

With three active flavours of quarks [8], then in the $\overline{MS}$ renormalization scheme [29] explicit calculation results in 
\begin{equation}\tag{42a-d}
b = 9/4, \quad c = 16/9, \quad c_2 = 3863/869, \quad c_3 = 20.99024031 
\end{equation}
where the values of $b$ and $c$ are the same in any mass--independent renormalization scheme while the values of $c_2$ and $c_3$ in eqs. (42c,d) are peculiar to the $\overline{MS}$ scheme.  Furthermore, we find in ref. [30] that in the $\overline{MS}$ scheme
\begin{equation}\tag{43a-d}
T_{0,0} = 1, \quad T_{1,0} = 1.6401, \quad T_{2,0} = -10.28395,\quad T_{3,0} = 20.99024031
\end{equation}
which by eq. (31) results in
\begin{equation}\tag{44a-c}
\tau_1 = 1.6401, \quad \tau_2 = -5.812885185 \quad \rm{and}  \quad \tau_3 = -81.73499303.
\end{equation}
At the $\tau$ mass $( m_\tau = 1.777 GeV/c^2)$, using the $\overline{MS}$ scheme, we have [45] (with $a(\mu)$ defined by eq. (2)),
\begin{equation}\tag{45}
a_{\overline{MS}} (\mu = m_\tau ) = \frac{.33}{\pi}.
\end{equation}
We have arbitrarily chosen the value of $\mu$ at which $a_{\overline{MS}}(\mu)$ is evaluated to be $m_\tau$, but by use of eq. (24) it is possible to convert this value of $a_{\overline{MS}}$ to the one corresponding to other values of $\mu$.

On fig. 1 we first plot the perturbative result from eq. (1)
\begin{equation}\tag{
}
1 + R_{\rm{pert}}^{IV}  = 1 + \sum_{n=0}^3 \sum_{m=0}^n T_{n,m} \left( b \ln \left(\frac{m_\tau}{Q}\right)\right)^m a^{n+1}_{\scriptscriptstyle{\overline{MS}}} (m_\tau) 
\end{equation}
using the $\overline{MS}$ results of eqs. (42,43.45).  We also use eqs. (8,9) to plot the RG$\Sigma$ results in the $\overline{MS}$ scheme
\begin{equation}\tag{47}
1 + R_\Sigma^{IV} = 1 + \sum_{n=0}^3 a^{n+1}_{\overline{MS}} (m_\tau) S_n
\left( b a_{\overline{MS}} (m_\tau) \ln \left(\frac{m_\tau}{Q}\right)\right).
\end{equation}
on fig. 1.

Next we plot, from eq. (36), 
\begin{equation}\tag{48}
1 + R_A^{(1)IV} = 1 + a_{(1)}^{IV} \left( \ln \frac{Q}{m_\tau} \right)+ \tau_1 a_{(1)}^{IV\,2} \left( \ln \frac{Q}{m_\tau} \right)
\end{equation}
on fig. 1.  In order to obtain $a_{(1)}^{IV}$ in eq. (48), we first use eq. (34) to obtain $a_{I(1)} (m_\tau)$ (eq. (45)) and then employ eq. (25) to obtain 
 $a_{(1)}^{IV} \left( \ln \frac{Q}{m_\tau} \right)$ from $a_{I(1)}$. (We have taken $\Lambda$ to be $m_\tau$ in eq. (14).)
 
Finally on fig. 1 we have plotted
\begin{equation}\tag{49}
1 + R_A^{(2)IV} = 1 + a_{(2)}^{IV} \left( \ln \frac{Q}{m_\tau} \right)+ \tau_1 a_{(2)}^{IV\,2} \left( \ln \frac{Q}{m_\tau} \right) 
+ \tau_2 a_{(2)}^{IV\,3} \left( \ln \frac{Q}{m_\tau} \right) + \tau_3 a^{(IV)4}_{(2)} \left( \ln \frac{Q}{m_\tau} \right). 
\end{equation}
In eq. (49), $a_{(2)}^{IV} \left( \ln \frac{Q}{m_\tau} \right)$ is obtained from
$a_{(1)}^{IV} \left( \ln \frac{Q}{m_\tau} \right)$ using eq. (34).

In $R_{\rm{pert}}$ (eq. (1)) and $R_\Sigma$ (eq. (9)), there is an explicit $\mu$ dependence; in eq. (46,47) we have chosen $\mu$ to equal $m_\tau$.  However, all $\mu$ dependence has disappeared in the alternatively summed $R_A$ in eq. (18); one only has a scale parameter $\Lambda$ that is the mass scale at which the boundary condition of eq. (2) is fixed (see. eq. (14)).  We have chosen, for convenience, to set $\Lambda = m_\tau$.  Consequently, $R_A^{(1)IV}$ and $R_A^{(2)IV}$ are independent of $\mu$.  We note that $R_A^{(1)IV}$ and $R_A^{(2)IV}$ almost coincide over the entire range of $Q$, and neither are affected by the occurrence of large logarithms when $Q$ is far from $m_\tau$, where as $R^{IV}_{\rm{pert}}$ and  $R^{IV}_\Sigma$ both exhibit a strong dependence on large logarithms far from $m_\tau$.

In fig. 2 we plot $a_{(1)}^{IV} \left( \ln \frac{Q}{m_\tau} \right)$ and 
$a_{(2)}^{IV} \left( \ln \frac{Q}{m_\tau} \right)$ as functions of $Q$, again for $.5 \leq Q \leq 4 (GeV/c^2)$. Both exhibit asymptotic freedom, but they are quite distinct.  Never the less, the value of $1 + R_A^{(1) IV}$ and  $1 + R_A^{(2) IV}$ are nearly coincident in this range.

In fig. 3, $Q$ is fixed at $m_\tau$ and a plot of $1 + R_{\rm{pert}}^{IV}$ and  $1 + R_\Sigma^{IV}$ against $\mu$ is presented.  This illustrates how the $RG\Sigma$ procedure reduces $\mu$ dependency, but does not eliminate it.

\section{Discussion}

Possibly the most interesting aspect of the approach to determining $R$ from perturbative calculations that we have outlined is the absence of dependence of our result on the renormalization scale parameter $\mu$. One can choose any convenient value of $\Lambda$; the value of $Q$ is some multiple of this value and the value of the running coupling $a$ at this value of $\Lambda$ is determined by experiment. 

We have also shown how all coefficients $T_{nm}$ in eq. (1) that are computed by a perturbative calculation can be expressed in terms of the coefficient $b$, $c$, $c_i$ appearing in the perturbative expansion of the function $\beta$ of eq. (2) and a set of renormalization scheme invariants $\tau_i$ occurring in eq. (31).  These $\tau_i$ are related to the renormalization scheme invariants $\rho_i$ previously found by Stevenson [13,31].  For example, the first scheme invariant of Stevenson is 
\begin{subequations}
\begin{align}
\rho_2 &= L - r_1 \nonumber \\
&= T_{00}L - (T_{10} + T_{11}L).\tag{50a}\\
\intertext{By eq. (10a) and (31) we find that}
\rho_2 &= -\tau_1.\tag{50b}
\end{align}
\end{subequations}
Next, the invariant
\begin{equation}\tag{51}
\rho_3 = c_3 + 2r_3 - 2c_2 r_1 - 6 r_2r_1 + c r_1^2 + 4r_1^3
\end{equation}
with $r_n = \displaystyle{\sum_{m=0}^n} T_{n,m}L^m$ is satisfied on account of eqs. (10,31) provided
\begin{equation}\tag{52}
\rho_3 =  2\tau_3 -  6 \tau_2\tau_1 + c \tau_1^2 + 4\tau_1^3 ;
\end{equation}
with no dependency on the scheme dependent quantities $c_i$.  In general $\rho_n$ will depend on $c, \tau_1 \ldots \tau_N$ but be independent of $L, c_i (i \geq 2)$.

In scheme (2) where the $\beta$ function consists of two terms, it is well known that a ``infrared fixed point'' occurs when $\beta(a) = 0$; this means that by eq. (38) [33] when $a$ reaches a value that satisfies
\begin{equation}\tag{53}
1 + ca = 0
\end{equation}
the running coupling no longer evolves.  Acceptable values for $a$ must be positive; furthermore $b > 0$ in order for the theory to be asymptotically free.  As $b = \frac{33-2N_f}{6}$, $c = \frac{153-19N_f}{2(33-2N_f)}$ [33] in massless QCD with $N_f$ flavours, asymptotic freedom $(b > 0)$ and a positive solution for $a$ in eq. (53) occur if 
\begin{subequations}
\begin{align}
N_f & \leq \frac{33}{2} \approx 16 \tag{54a}\\
\intertext{and}
N_f & \geq \frac{153}{19} \approx 8\tag{54b}.
\end{align}
\end{subequations}
The bounds on $N_f$ in eq. (53) are outside the physically realized range of values for the number of quark flavours. This was realized by Caswell and Jones who first computed $c$ [33] and more extensively discussed by Banks and Zaks [33].  However, if we were to adopt the $\beta$ function associated with $a_{(3)}$ in eq. (41), then an infrared fixed point occurs if 
\begin{subequations}
\begin{align}
 1 & + ca + c_2a^2 = 0 \tag{55a}\\
\intertext{or}
a & = \frac{+1 \pm \sqrt{1 + \frac{4\lambda}{c}}}{2\lambda} \tag{55b}
\end{align}
\end{subequations}
where $c_2 = -\lambda c$.  If $N_f = 3$, then $c = \frac{48}{27}$ and so upon using the positive root in eq. (55b)
\begin{equation}\tag{56}
a = \frac{1 + \sqrt{1+27\lambda/12}}{2\lambda}
\end{equation}
which decreases for $\lambda > 0$ as $\lambda$ grows, and is less than one for $\lambda \gtrsim 1.6$.  Thus when using this renormalization scheme, an infrared fixed point occurs when there is a physically realizable number of quark flavours, $N_F = 3$. 

If two renormalization schemes lead to the couplings $a$, $\overline{a}$ related by eq. (3a), the $a$ and $\overline{a}$ evolve under changes in $\ln \mu$ with distinct $\beta$ functions. An infrared fixed point occurs at a zero of the $\beta$ function and hence it is apparent that the location of such a fixed point is renormalization scheme dependent.  Differentiating eq. (3a) with respect to $\ln \mu$ leads to 
\begin{equation}\tag{57}
\mu \frac{d\overline{a}}{d\mu} \equiv \overline{\beta} (\overline{a}) = \beta (a) F^\prime (a).
\end{equation}
If $a^*$ satisfies $\beta(a^*) = 0$ (for example, by satisfying eq. (53) when dealing with $a_{(2)}$), then it doesn't follow that $\overline{\beta}(\overline{a}^*) = 0$ if $\overline{a}^* = F(a^*)$ (for example, if $\overline{a} = a_{(3)}$).  From eq. (57), this means that $F^\prime (a^*)$ diverges.

It is often stated that although the location of a fixed point is renormalization scheme dependent [38,39] the slope of the $\beta$ function at a fixed point is scheme independent. The argument used is that 
\begin{equation}\tag{58}
\frac{\overline{\beta}(\overline{a})}{\beta(a)} = \frac{\mu \frac{d\overline{a}}{d\mu}}{\mu \frac{d\overline{a}}{d\mu}} = \frac{d\overline{a}}{da}
\end{equation}
and so by eq. (57)
\begin{subequations}
\begin{align}
\frac{d\overline{\beta}(\overline{a})}{d\overline{a}} &= \frac{1}{F^\prime(a)} \frac{d}{da} \left( \beta (a) F^\prime (a)\right)\nonumber \\
& = \frac{d\beta(a)}{da} + \beta (a) \left (\frac{d^2F}{da^2}/ \frac{dF}{da} \right).\tag{59}
\end{align}
\end{subequations}
Having $\beta(a^*) = 0$ leads to $\overline{\beta}^\prime (\overline{a}^*) = 0$ when 
$\beta^\prime (a^*) = 0$ only if $F^{\prime\prime} (a^*)/F^\prime(a^*)$ is finite but we have seen from eq. (57) that this need not be the case. This has been further discussed in ref. [40] and it was demonstrated using Pad$\rm{\acute{e}}$ summation techniques that no infrared fixed point exists for low $n_f$ in the $\overline{MS}$ scheme [41].  On the contrary, a majority of lattice QCD results have found that while such a fixed point exists, it occurs in most studies with $n_f = 12$ flavours [42]. (For an overview of various lattice results, see also ref. [43].) It is also worthwhile to note that it was recently demonstrated in [44] that via two new one-parameter families of scheme transformations, for moderate values of gauge coupling and parameters specifying the scheme transformation, the effect of scheme dependence on the infrared fixed point was found to be mild. 

If in eq. (3a), $a = a_{(2)}$ and $\overline{a} = a_{(3)}$ then $c_i = 0 ( i \geq 2), \overline{c}_i = 0 ( i \geq 3)$ and eq. (3b) becomes
\begin{equation}\tag{60}
x_3 = \overline{c}_2 = cx_2 + x_2^2
\end{equation}
while by eq. (3c)
\begin{equation}\tag{61}
x_4 = \frac{1}{2} \left[ 3x_2^2 c + 4\overline{c}_2x_2 + 2x_2x_3\right]
\end{equation}
etc.\\
with $x_2$ not being specified.  Consequently for a given value of $a = a_{(2)}$, there corresponds a continuum of values of $\overline{a} = a_{(3)}$.  In general we have $x_n (n > 2)$ determined by $c$, $\overline{c}_2$ and $x_2$.  In ref. [13] it is argued that when parameterizing renormalization scheme dependence, $x_2$ can be identified with $\mu$ while $x_3, x_4 \ldots$ are associated with $c_2, c_3 \ldots$.  We have found a way of using $RG$ summation to eliminate dependence on the scale parameter $\mu$ (see eq. (18)) in $R$ and so $x_2$ is a free parameter.  Thus eq. (57), which is the equation used to arrive at eqs. (3b, 3c), does not fix $\overline{a}$ uniquely for a given value of $a$ when $\beta(a)$ and $\overline{\beta}(\overline{a})$ are chosen at the outset (for example, $\beta(a) = -ba^2(1 + ca)$ and $\overline{\beta}(\overline{a}) = -b\overline{a}^2 (1 + c \overline{a} + \overline{c}_2 \overline{a}^2)$).

It is interesting to note that in the scheme in which the perturbative expansion for $R$ terminates after two terms (see eq. (36)), the question of the existence of ``renormalons'' [26] gets shifted to an analysis of the function $a_{(1)} \left( \ln \frac{Q}{\Lambda} \right)$.  This is because renormalons arise in the discussion of the asymptotic behaviour of the infinite series in which $R$ is expanded in powers of the coupling; with the choice of renormalization scheme made for $a_{(1)}$, the expansion for $R$ (eq. (36)) is no longer an infinite series in powers of the coupling $a_{(1)}$ but terminates after two terms. Higher loop computations when using this scheme serve to determine the renormalization scheme invariants $\tau_i$ associated with this process which in turn fix the values of $c_i (i \geq 2)$ in this scheme (eq. (35)).

We are currently examining the extension of the ideas presented in this paper to situations in which massive fields or multiple couplings occur.  We would also like to consider scheme and gauge dependence when using mass dependent renormalization schemes such as $MOM$\vspace{.52cm}.
\\
\noindent
{\Large\bf{Acknowledgements}}\\
R. Macleod had helpful input.

\begin{figure}[hbt]
\begin{center}
\includegraphics[scale=0.9]{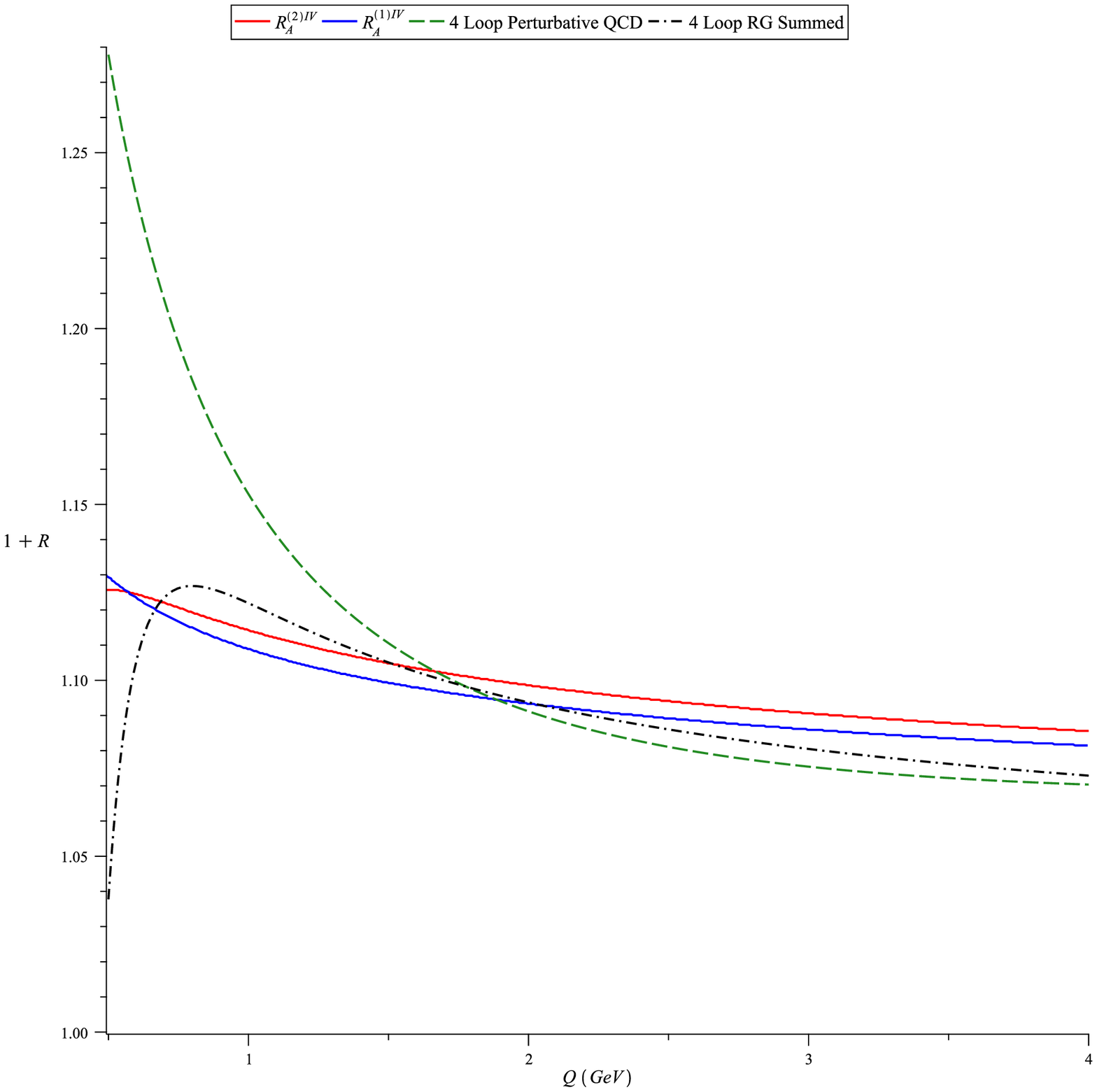}
\caption{The $Q$ dependence of renormalization scale independent $1 + R_A^{(1) IV}$ and  $1 + R_A^{(2) IV}$ at $\Lambda=m_{\tau}$ as compared to 4-Loop Perturbative QCD and 4-Loop RG summed results for $1+R$ referenced at renormalization scale, $\mu=m_{\tau}$}
\label{Fig. 1}
\end{center}
\end{figure}

\begin{figure}[hbt]
\begin{center}
\includegraphics[scale=0.9]{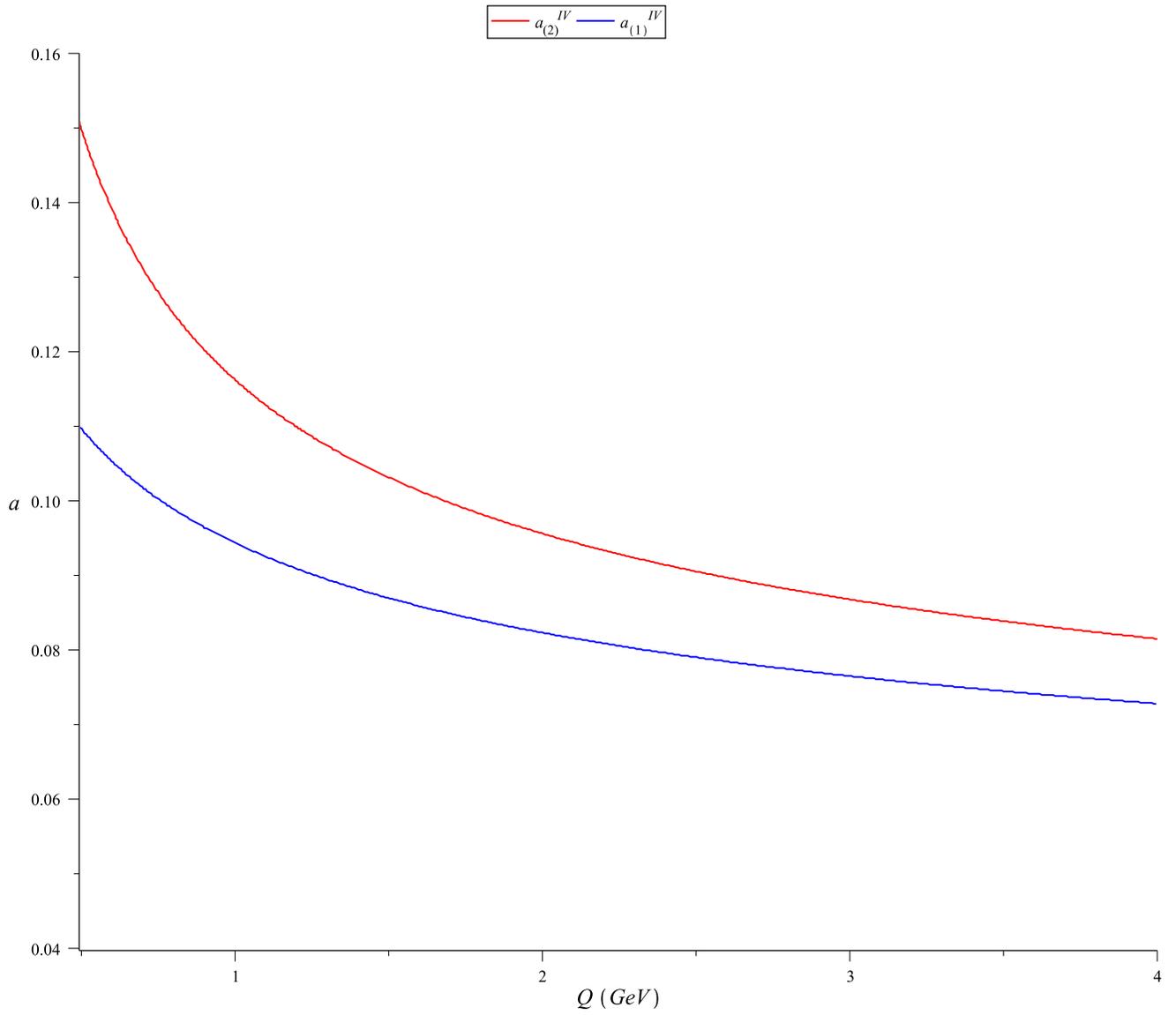}
\caption{The $Q$ dependence of renormalization scale independent $a_{(1)}^{IV}$  and $a_{(2)}^{IV} $}
\label{Fig. 2}
\end{center}
\end{figure}

\begin{figure}[hbt]
\begin{center}
\includegraphics[scale=0.9]{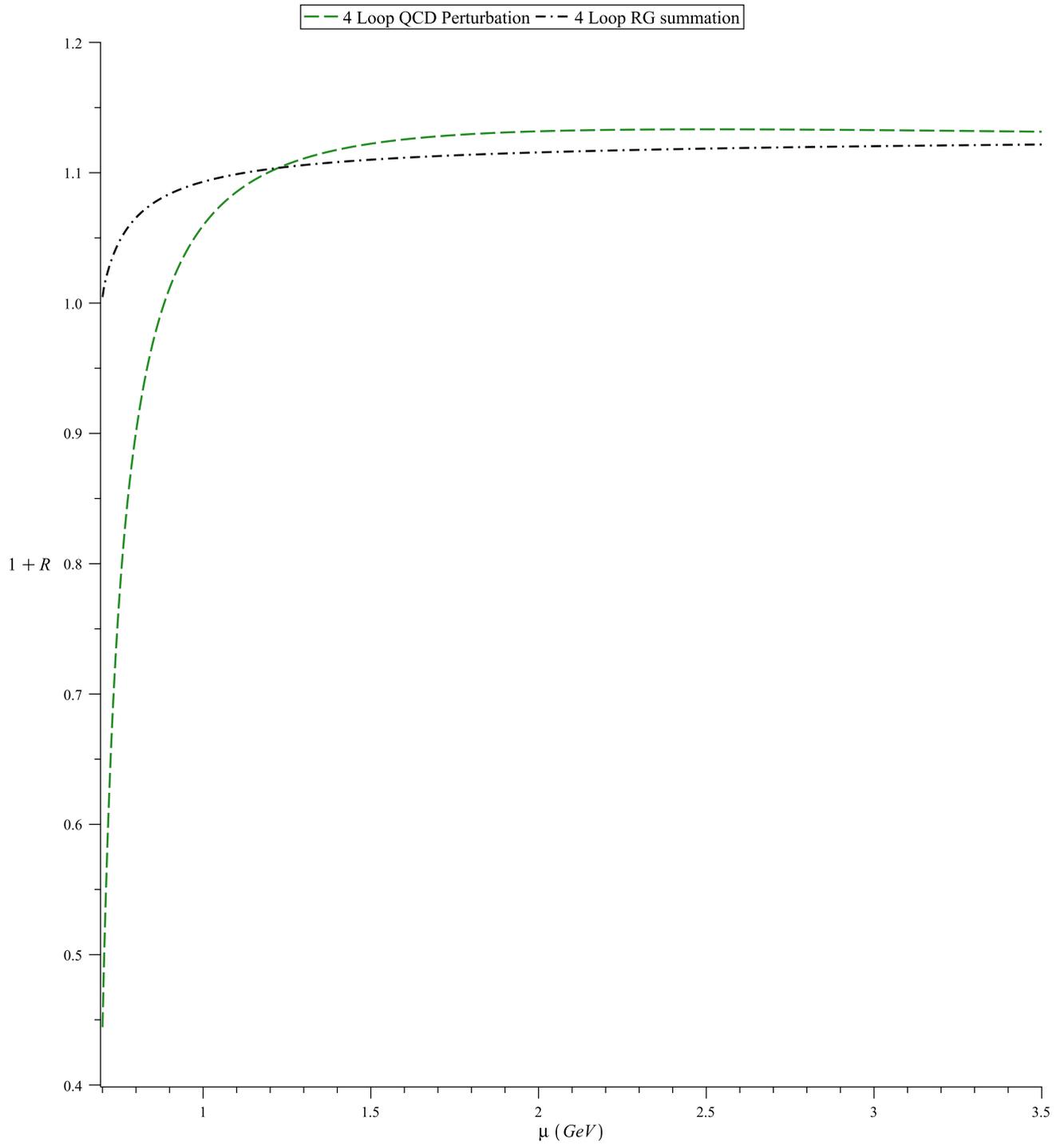}
\caption{The renormalization scale $\mu$ dependence of $1+R$ for 4-Loop Perturbative QCD and 4-Loop RG summed results when $Q=m_{\tau}$}
\label{Fig. 3}
\end{center}
\end{figure}

\end{document}